
\magnification=1200
\baselineskip=13pt
\overfullrule=0pt
\tolerance=100000
\rightline{UR-1362\ \ \ \ \ \ \ }
\rightline{ER-40685-812}

\bigskip

\baselineskip=18pt

\centerline{\bf THE SUPERSYMMETRIC TWO BOSON
HIERARCHIES}

\bigskip
\bigskip

\centerline{J.C. Brunelli}
\centerline{and}
\centerline{Ashok Das}
\centerline{Department of Physics}
\centerline{University of Rochester}
\centerline{Rochester, NY 14627}

\vskip 1.7 truein

\centerline{\bf \underbar{Abstract}}

\medskip

We construct the most general supersymmetric two boson system that
is integrable. We obtain the Lax operator and the nonstandard Lax
representation for this system. We show that, under appropriate
redefinition of variables, this reduces to the supersymmetric nonlinear
Schr\"odinger equation without any arbitrary parameter which is known
to be integrable. We show that this supersymmetric system has three local
Hamiltonian structures just like the bosonic counterpart and we show
how the supersymmetric KdV equation can be embedded into this system.

\vfill\eject

\noindent {\bf I. \underbar{Introduction}:}

\medskip

Integrable systems in 1+1 and 2+1 dimensions have been studied vigorously
in the past [1-3]. These are nonlinear systems with a biHamiltonian structure
and with infinitely many, independent, commuting conserved quantities
and possess solitonic solutions. Furthermore, these equations have a
standard Lax representation which makes them integrable through the
method of inverse scattering. Such systems have also appeared in various
studies in string theory [4].

More recently, a dispersive generalization of the long water wave
equation [5-7] has received much attention [8-11]. It has the form
$$\eqalign{{\partial u \over \partial  t} &= (2h + u^2 - \alpha
 u^\prime)^\prime\cr
\noalign{\vskip 4pt}%
{\partial  h \over \partial  t} &= (2uh + \alpha h^\prime)^\prime\cr}\eqno(1)$$
where $u(x,t)$ and $h(x,t)$ can be thought of as the horizontal velocity
and the height, respectively,
 of the free surface ($\alpha$ is  an arbitrary parameter)
and a prime denotes a derivative with respect to $x$. The system of
equations in (1) is integrable [6] and has a triHamiltonian structure [7].
 It has
a nonstandard Lax representation and reduces to various known integrable
systems with appropriate identification [7]. Thus, for example, with the
identification
$$\eqalign{\alpha &= 1\cr
u &= - q^\prime / q\cr
h &= q \overline  q\cr}\eqno(2)$$
equations (1) reduce to the nonlinear Schr\"odinger equation [8-11].
 With $\alpha =-1$,
$h=0$, Eq. (1) gives the Burgers' equation while for $\alpha=0$ we obtain
Benney's equation from (1). For the rest of our discussion, we will choose
$\alpha = 1$.

While the supersymmetric nonlinear Schr\"odinger equation has been studied
in some detail in recent years [12-14]
 and while a fermionic extension of Eq. (1)
has also been investigated [15], the supersymmetric form of Eq. (1) has not yet
been obtained. The study of such a system, among other things, is expected
to shed light on such properties as superturbulence in classical
 hydrodynamics.
 In this paper, we study the supersymmetric generalization of the
long water wave equation. In sec. II, we obtain the most general superfield
equation consistent with dimensional analysis which can also be expressed
as a superfield Lax equation. We give the component form of the equations
and the supersymmetry transformations which leave the system invariant. In
sec. III, we show how this equation reduces to the supersymmetric nonlinear
Schr\"odinger equation which is integrable and which has been studied in some
detail [13]. In sec. IV, we derive the Hamiltonian structures for this system
and show that it is a triHamiltonian system much like the bosonic equation
in (1). Finally, in sec. V we show how other supersymmetric equations,
such as the susy KdV, can be  embedded into this system and present a
brief conclusion in sec. VI.

\medskip

\noindent {\bf II. \underbar{Supersymmetric Equation}:}

\medskip

For the rest of our discussions, we will choose $\alpha=1$ and for consistency
with the two boson formulation of this system [8,16], we
will make the identification
$u=J_0$ and $h=J_1$. The equations in (1) then take the form
$$\eqalign{{\partial  J_0 \over \partial  t} &= (2 J_1 + J_0^2 -
J^\prime_0 )^\prime\cr
\noalign{\vskip 4pt}%
{\partial  J_1 \over \partial  t} &= (2 J_0 J_1 +
J^\prime_1 )^\prime\cr}\eqno(3)$$

A simple dimensional analysis shows that we can assign the following
canonical dimensions to the variables of the system.
$$[x] = -1 \qquad [t] =-2 \qquad [J_0] = 1 \qquad [J_1 ] = 2 \eqno(4)$$
This system can be represented as a Lax equation [7,16]
$${\partial  L \over \partial  t} = \left[ L, \left( L^2 \right)_{\geq 1}
\right] \eqno(5)$$
where
$$L =  \partial  - J_0 + \partial^{-1} J_1 \eqno(6)$$
and $()_{\geq 1}$ refers to the differential part of a
pseudo-differential operator. (For details see ref. 16) We note that
$$[L] = 1\eqno(7)$$

The simplest way to obtain the supersymmetric equation is to go to the
superspace. Let $z = (x,\theta)$ define the superspace and
$$D = {\partial  \over \partial  \theta} + \theta {\partial
 \over \partial  x}\eqno(8)$$
represent the supercovariant derivative. From the relation $D^2 = \partial$,
it follows that
$$[\theta] = - {1 \over 2}\eqno(9)$$
Let us next introduce two fermionic superfields
$$\eqalign{\Phi_0 &= \psi_0 + \theta J_0\cr
\Phi_1 &= \psi_1 + \theta J_1 \cr}\eqno(10)$$
The canonical dimensions of the new variables now follow to be
$$\eqalign{
[\Phi_0] &= [ \psi_0] = {1 \over 2}\cr
\noalign{\vskip 4pt}%
[\Phi_1] &= [ \psi_1] = {3 \over 2}\cr}\eqno(11)$$

Given these, one can write the most general dynamical equations in superspace
consistent with the canonical dimensions and which reduce to Eq. (3) in
the bosonic limit as
$$\eqalign{{\partial  \Phi_0 \over \partial  t} = &-(D^4 \Phi_0) + 2(D \Phi_0)
(D^2 \Phi_0) + 2(D^2 \Phi_1)\cr
&+ a_1 D(\Phi_0 (D^2 \Phi_0)) + a_2 D (\Phi_0 \Phi_1)\cr
\noalign{\vskip 4pt}%
{\partial  \Phi_1 \over \partial  t} = &(D^4 \Phi_1) + b_1 D( ( D^2
\Phi_1 ) \Phi_0) + 2(D^2 \Phi_1)(D \Phi_0 ) - b_2 D(\Phi_1
(D^2 \Phi_0))\cr
& +2(D \Phi_1 )(D^2 \Phi_0) + b_3 \Phi_1 \Phi_0(D^2 \Phi_0 ) + b_4 D(\Phi_1
\Phi_0) (D \Phi_0)\cr
&+ b_5 D(\Phi_0 (D^4 \Phi_0)) +b_6 D(\Phi_0 (D^2 \Phi_0)) (D \Phi_0)\cr}
\eqno(12)$$
Here $a_1$, $a_2$, $b_1$, $b_2$, $b_3$, $b_4$, $b_5$ and $b_6$ are
arbitrary parameters and Eq. (12) represents the most general supersymmetric
extension of Eq. (3). This system of equations, however, maynot be
integrable. To find an integrable, supersymmetric extension, we look for a Lax
representation for the system of equations. We find that a consistent Lax
representation can be obtained if we define
$$L = D^2 + \alpha (D \Phi_0) + \beta D^{-1} \Phi_1 \eqno(13)$$
where $\alpha$, $\beta$ are arbitrary parameters. (Note that for $\alpha=-1$
and $\beta = 1$ Eq. (13) reduces to Eq. (6) in the bosonic limit.)
 The nonstandard
Lax equation
$${\partial  L \over \partial  t} = \left[ L, (L^2)_{\geq 1}
\right]\eqno(14)$$
in this case, gives
$$\eqalign{{\partial  \Phi_0 \over \partial  t} &= -(D^4 \Phi_0) - 2 \alpha (D
\Phi_0)
(D^2 \Phi_0) - {2 \beta \over \alpha} \ (D^2 \Phi_1)\cr
\noalign{\vskip 4pt}%
{\partial  \Phi_1 \over \partial  t} &= (D^4 \Phi_1) - 2 \alpha D^2
((D \Phi_0 )\Phi_1)\cr}\eqno(15)$$
Comparing with Eq. (12), we conclude that $\alpha = -1$ and
 $\beta = 1$ so that
$$L = D^2 - (D \Phi_0) + D^{-1} \Phi_1\eqno(16)$$
and it would appear that the most general supersymmetric extension of Eq. (3)
which is integrable is given by
$$\eqalign{{\partial  \Phi_0 \over \partial  t} &=
 - (D^4 \Phi_0) + 2 (D \Phi_0)(D^2 \Phi_0)
                     + 2(D^2 \Phi_1)\cr
\noalign{\vskip 4pt}%
{\partial  \Phi_1 \over \partial  t} &=
 (D^4 \Phi_1) + 2 D^2((D \Phi_0) \Phi_1)\cr}\eqno(17)$$
In components, the equations have the form
$$\eqalign{
{\partial  J_0 \over \partial  t} &= ( 2J_1 + J_0^2 - J_0^\prime)^\prime\cr
\noalign{\vskip 4pt}%
{\partial  \psi_0 \over \partial  t} &= 2 \psi_1^\prime + 2 \psi_0^\prime
J_0 - \psi_0^{\prime \prime}\cr
\noalign{\vskip 4pt}%
{\partial  J_1 \over \partial  t} &= ( 2J_0 J_1 + J_1^\prime
 + 2 \psi_0^\prime \psi_1)^\prime\cr
\noalign{\vskip 4pt}%
{\partial  \psi_1 \over \partial  t} &= ( 2 \psi_1
J_0 + \psi_1^\prime)^\prime\cr}\eqno(18)$$
This is a completely interacting system and is invariant under the
supersymmetric transformations
$$\eqalign{
\delta \psi_0 &= \epsilon J_0\cr
\delta  J_0 &= \epsilon \psi_0^\prime\cr
\delta  \psi_1 &= \epsilon J_1\cr
\delta  J_1 &= \epsilon \psi_1^\prime\cr}\eqno(19)$$
where $\epsilon$ is a constant Grassmann parameter of transformation.

\medskip

\noindent {\bf III. \underbar{Reduction to SUSY NLS Equation}:}

\medskip

As we have noted in the introduction, the bosonic equation (3) reduces to
the nonlinear Schr\"odinger equation with appropriate redefinition of
variables (see Eq. (2)). It is, therefore, natural to expect that the
supersymmetric equations (17) will reduce to the supersymmetric nonlinear
Schr\"odinger equation [12-14]
 with appropriate redefinition of variables. This is
indeed true and it is worth emphasizing that the only consistent set of
field redefinitions that is possible, leads to the supersymmetric
nonlinear Schr\"odinger equation without any arbitrary parameter. Without
going into detail, we note that the field redefinitions
$$\eqalign{
\Phi_0 &= - D \ln (DQ) +
D^{-1} (\overline Q Q)\cr
\noalign{\vskip 4pt}%
\Phi_1 &= - \overline Q ( DQ)\cr}\eqno(20)$$
where $Q = \psi + \theta q$ and $\overline Q = \overline \psi + \theta
\overline q$
 are fermionic superfields which are complex conjugates
of each other, leads from Eq. (17) to (The derivation is slightly involved.)
$$\eqalign{{\partial  Q \over \partial  t} &=
 -(D^4 Q) + 2((D^2 Q) \overline Q +(DQ)(D \overline Q ))Q\cr
\noalign{\vskip 4pt}%
{\partial  \overline Q \over \partial  t} &=
 (D^4 \overline Q ) - 2((D^2 \overline Q ) Q + (D \overline Q)(DQ))
\overline Q\cr}\eqno(21)$$

These are nothing other than the supersymmetric nonlinear Schr\"odinger
equations without any free parameter. (See ref. 13  for details.)
 These equations
are known to be integrable which also proves integrability of Eq. (17). We
note that the field redefinitions in Eq. (20) reduce to Eq. (2) (with
appropriate identification) in the bosonic limit.

\medskip

\noindent {\bf IV. \underbar{Hamiltonian Structures}:}

\medskip

Most bosonic integrable systems are known to be biHamiltonian. The
corresponding supersymmetric systems, on the other hand, have only a
single Hamiltonian structure which is local (see [17-19] for the
susy KdV system). As we have indicated in
the introduction, the present bosonic integrable system in Eq. (3) is
a triHamiltonian system [7]. It is, therefore, interesting to ask what will be
the Hamiltonian structures of the supersymmetric equation in Eq. (17). We
find that the present supersymmetric system is also triHamiltonian
and this implies its integrability.

It is easy to verify with the usual Berezin rules for Grassmann variables
that with $( z= (x,\theta))$
$$\eqalign{
\{\Phi_0 (z_1),\Phi_0(z_2)\}_1 &= 0 = \{\Phi_1(z_1),\Phi_1(z_2)\}_1\cr
\{\Phi_0(z_1),\Phi_1(z_2)\}_1 &= D_{z_{2}} \partial  (z_1 -z_2)
                              = \{\Phi_1(z_1),\Phi_0(z_2)\}_1\cr}\eqno(22)$$
and with
$$\eqalign{H_1 = &\int
 dz \big[ -\Phi_1(z) (D \Phi_1 (z)) + (D^3 \Phi_0 (z)) \Phi_1 (z)\cr
\noalign{\vskip 4pt}%
                   &- \Phi_0 (z) (D \Phi_0 (z)) (D \Phi_1 (z))
                   - (D^2 \Phi_0 (z))\Phi_1 (z) \Phi_0 (z)\big]\cr}\eqno(23)$$
we can write Eq. (17) in the Hamiltonian form
$$\eqalign{
{\partial  \Phi_0 (z) \over \partial  t} &= \{\Phi_0(z), H_1\}_1\cr
\noalign{\vskip 4pt}%
{\partial  \Phi_1  (z) \over \partial  t} &=\
 \{\Phi_1 (z), H_1\}_1\cr}\eqno(24)$$
On the other hand, we can choose
$$\eqalign{
\{\Phi_0(z_1), \Phi_0(z_2)\}_2 &= 2D_{z_{2}} \partial  (z_1 -z_2)\cr
\{\Phi_0 (z_1),\Phi_1 (z_2)\}_2 &= (D_{z_{2}} \Phi_0 (z_2))
D_{z_{2}} \partial  (z_1 - z_2) +
   D_{z_{2}}^3 \partial  (z_1 - z_2)\cr
\{\Phi_1 (z_1),\Phi_0 (z_2)\}_2 &= (D_{z_{1}} \Phi_0 (z_1))
D_{z_{2}} \partial  (z_1 - z_2) -
     D_{z_{2}}^3 \partial  (z_1 - z_2)\cr
\{\Phi_1 (z_1), \Phi_1 (z_2)\}_2 &= 2 \Phi_1 (z_1) D_{z_{2}}^2
\partial  (z_1 - z_2) - (D_{z_{2}}^2 \Phi_1 (z_2))
\partial  (z_1 - z_2)\cr}\eqno(25)$$
and
$$H_2 = -\int dz \Phi_1 (z) (D \Phi_0 (z))\eqno(26)$$
to write  Eq. (17) also in the Hamiltonian form
$$\eqalign{
{\partial  \Phi_0 (z)\over \partial  t} &= \{ \Phi_0 (z), H_2\}_2\cr
\noalign{\vskip 4pt}%
{\partial  \Phi_1 (z) \over \partial  t} &= \{ \Phi_1 (z), H_2\}_2\cr}
\eqno(27)$$

Finally, it is also easy to verify that the choice
$$
\eqalign{
\{\Phi_0 (z_1), \Phi_0 (z_2)\}_3 =&
2\big[ \big( D_{z_1} \Phi_0 (z_1)\big) + \big( D_{z_2} \Phi_0 (z_2)\big)
\big] D_{z_2} \delta (z_1 - z_2)\cr
\{\Phi_0 (z_1), \Phi_1 (z_2)\}_3 =& \big[ D^2_{z_2} + \big( D_{z_2}
\Phi_0 (z_2) \big) \big]^2 D_{z_2} \delta (z_1 - z_2 )\cr
 &+ 2 \big(\Phi_1 (z_1)+ \Phi_1 (z_2) \big) D^2_{z_2}
\delta (z_1 - z_2)\cr
\{\Phi_1 (z_1), \Phi_0 (z_2 )\}_3 =& \big[ D^2_{z_1} +  \big( D_{z_1}
\Phi_0 (z_1) \big) \big]^2 D_{z_2} \delta (z_1 - z_2)\cr
 &+2 \big( \Phi_1 (z_1)+ \Phi_1 (z_2) \big)
D^2_{z_2}\delta (z_1 - z_2)\cr
\{\Phi_1 (z_1), \Phi_1 (z_2)\}_3 =&\big[ D^2_{z_1} + \big( D_{z_1} \Phi_0
 (z_1)\big)\big]\big( \Phi_1 (z_1) + \Phi_1 (z_2)
\big) D^2_{z_2} \delta (z_1 - z_2)\cr
+&\big[ D^2_{z_2} + \big( D_{z_2} \Phi_0 (z_2) \big) \big]
\big( \Phi_1 (z_1) + \Phi_1 (z_2) \big) D^2_{z_2} \delta (z_1 - z_2)\cr}
\eqno(28)
$$
and the Hamiltonian
$$H_3 = - \int dz \Phi_1(z)\eqno(29)$$
would also make Eq. (17) Hamiltonian. Thus, unlike other supersymmetric
integrable systems, the present system allows generalization of all the
three Hamiltonian structures of its bosonic counterpart. Let us note that
if we denote the three Hamiltonian structures of Eqs. (22), (25) and (28) in
the matrix form as $\omega_1^{-1}$, $\omega_2^{-1}$ and $\omega_3^{-1}$
respectively, then we can write
$$\eqalign{
\omega_2^{-1} &= R\, \omega_1^{-1}\cr
\omega_3^{-1} &= R\, \omega_2^{-1} = R^2 \omega_1^{-1}\cr}\eqno(30)$$
where
$$R = \omega_2^{-1} \omega_1\eqno(31)$$
This implies that the corresponding symplectic forms are also related
through a matrix operator
$$S = R^{-1}\eqno(32)$$
This immediately implies that the Nijenhuis tensor associated with
$S$ must vanish [20]which is also a sufficient condition for the integrability
of the system [21].

\medskip

\noindent {\bf V. \underbar{Embedding of SUSY KdV}:}

\medskip

It is known that the KdV equation can be embedded into the long water
wave equation and can also be expressed in the nonstandard Lax
representation [7]. We show here that the supersymmetric KdV equation [17] can,
similarly, be embedded into the supersymmetric extension of the
long water wave equation or the supersymmetric two boson  hierarchy.

If we choose $\Phi_0 = 0$, then the Lax operator from Eq. (16) takes the
form
$$L = D^2 + D^{-1} \Phi_1\eqno(33)$$
We note that in this case,
$$(L^3)_{\geq 1} = D^6 + 3 D \Phi_1 D^2\eqno(34)$$
It can now be checked in a straightforward manner that the nonstandard
Lax equation
$${\partial   L \over \partial  t} = [ L, (L^3)_{\geq 1}]$$
in the present case yields
$${\partial  \Phi_1 \over \partial  t} =
 - (D^6 \Phi_1 ) - 3 D^2 (\Phi_1 (D \Phi_1))\eqno(35)$$
which we recognize to be the supersymmetric KdV equation [17]. This is the
embedding of the susy KdV equation in the nonstandard Lax representation.
As is well known, while KdV is a biHamiltonian system, susy KdV has only
one local Hamiltonian structure [17-19]. The embedding, therefore, is not
compatible with all the Hamiltonian structures of the system.

\medskip

\noindent {\bf VI.  \underbar{Conclusion}:}

\medskip

We have constructed the most general supersymmetric extension of the
two boson hierarchy (long water wave equation) which is integrable. We
have shown that this system reduces to the susy nonlinear Schr\"odinger
equation without any free parameter with appropriate redefinition of variables.
We have shown that this supersymmetric system is triHamiltonian much
like its bosonic counterpart. We have also shown how the supersymmetric
KdV equation can be embedded into this system. The derivation of
supersymmetric Burgers' equation and the supersymmetric Benney equation
can be carried out in a straight forward manner from our construction.

One of us (A.D.) would like to thank Prof. E. Witten and the Institute
for Advanced Study for their hospitality where part of this work was done.
This work was supported in part by the U.S. Department of Energy Grant No.
DE-FG-02-91ER40685.  J.C.B. would like to thank CNPq, Brazil, for
financial support.

\vfill\eject

\noindent {\bf \underbar{References}:}

\medskip

\item{1.} L.D. Faddeev and L.A. Takhtajan, \lq\lq Hamiltonian Methods in
the Theory of Solitons" (Springer, Berlin, 1987).

\item{2.} A. Das, \lq\lq Integrable Models" (World Scientific, Singapore,
1989).

\item{3.} M.J. Ablowitz and P.A. Clarkson, \lq\lq Solitons, Nonlinear
Evolution Equations and Inverse Scattering" (Cambridge, New York, 1991).

\item{4.} E. Nissimov and S. Pacheva, \lq\lq String Theory and Integrable
Systems", hep-th/9310113, and references therein.

\item{5.} L.J.F. Broer, Appl. Sci. Res. {\bf 31}, 377 (1975).

\item{6.} D.J. Kaup, Progr. Theor. Phys. {\bf 54}, 396 (1975).

\item{7.} B.A. Kupershmidt, Commun. Math. Phys. {\bf 99}, 51 (1985).

\item{8.} H. Aratyn, L.A. Ferreira, J.F. Gomes and A.H. Zimerman, Nucl.
 Phys. {\bf B402}, 85 (1993); H. Aratyn, L.A. Ferreira, J.F. Gomes and A.H.
Zimerman, \lq\lq $W_\infty$ Algebras, KP Hierarchies and its Gauge
Equivalences, Two-Boson Realizations and their KdV Reductions", hep-th/
9304152; H. Aratyn, E. Nissimov and S. Pacheva, Phys. Lett. {\bf B314},
41 (1993).

\item{9.} L. Bonora and C.S. Xiong, Phys. Lett. {\bf B285}, 191 (1992); L.
Bonora and C.S. Xiong, Int. J. Mod. Phys. {\bf A8}, 2973 (1993).

\item{10.} M. Freeman and P. West, Phys. Lett. {\bf 295B}, 59 (1992).

\item{11.} J. Schiff, \lq\lq The Nonlinear Schr\"odinger Equation and
Conserved Quantities in the Deformed Parafermion and SL(2,{\bf R})/U(1)
Coset Models", hep-th/9210029.

\item{12.} G.H.M. Roelofs and P.H.M. Kersten, J. Math. Phys. {\bf 33}, 2185
(1992).

\item{13.} J.C. Brunelli and A. Das, \lq\lq Tests of Integrability of the
Supersymmetric Nonlinear Schr\"odinger Equation", University of Rochester
preprint UR-1344 (1994) (also hep-th/9403019).

\item{14.} F. Toppan, ``$N$=1,2 Super-NLS Hierarchies as Super-KP Coset
Reductions", preprint ENSLAPP-L-467/94 (1994) (also hep-th/940595).

\item{15.} B.A. Kupershmidt, Mech. Res. Commun. {\bf 13}, 47 (1986).

\item{16.} J.C. Brunelli, A. Das and W.-J. Huang, \lq\lq Gelfand-Dikii
Brackets for Nonstandard Lax Equations", University of Rochester preprint
UR-1347 (1994) (also hep-th/940511).

\item{17.} P. Mathieu, J. Math. Phys. {\bf 29}, 2499 (1988).

\item{18.} A. Das and S. Roy, J. Math. Phys. {\bf 31}, 2145 (1990).

\item{19.} W. Oevel and Z. Popowicz, Comm. Math. Phys. {\bf 139}, 441
 (1991); J.M. Figueroa-O'Farril, J. Mas and E. Ramos, Leuven preprint
KUL-TF-91/19 (1991).

\item{20.} A. Das and W.-J. Huang, J. Math. Phys. {\bf 31} 2603 (1990); A.
Das, W.-J Huang and S. Roy, J. Math. Phys. {\bf 32}, 2733 (1991).

\item{21.} S. Okubo and A. Das, Phys. Lett. {\bf B209}, 311 (1988); A. Das
and S. Okubo, Ann. Phys. {\bf 190}, 215 (1989).

\end